# Analysis of the Frequency Offset Effect on Zadoff-Chu Sequence Timing Performance

Min Hua, Kristo W. Yang, M. Wang, Kingsley J. Zou

*Wireless Networking and Mobile Communications Group,*

*Nanjing University of Science and Technology*

*Abstract*—**Zadoff-Chu (ZC) sequences have been used as synchronization sequences in modern wireless communication systems, replacing the conventional pseudo-random noise (PN) sequences due to their perfect autocorrelation properties. We first study the ambiguity problem between a *timing offset* and a *frequency offset* when a ZC sequence is used as a synchronization signal. We show how a frequency offset can impair the timing property of a ZC sequence, causing irreducible timing errors. An analytical framework is developed, which completely characterizes a ZC sequence's timing behavior and its fundamental limitation as a synchronization sequence in the presence of a frequency offset between the transmitter and the receiver.**

## I. Introduction

SYNCHRONIZATION is the very first step in establishing a communication link to resolve the timing and frequency uncertainties between the transceivers, and is typically achieved by employing a special reference signal or pilot. For example, in a wireless communication system, an access point transmits a synchronization signal at a specific location of a transmission frame on the downlink. An access device acquires the system timing by searching for this special signal using a correlator matched to this signal. An access device then sends a signal to the access point on the uplink random access channel after aligning its local timing to the downlink synchronization signal in order to set up a connection with the network. The access point searches for the signal using a correlator, detecting the device's timing from the signal, and instructs the device to adjust its transmit timing to account for the round trip propagation delay before an uplink channel can be established on the assigned resources.

Signals providing good timing estimation must possess a good autocorrelation property [1]. Wireless communication systems almost unexceptionally employ sequences that have good autocorrelation properties to fulfill such a goal. In 2G and 3G systems (e.g., IS-95 [2], CDMA2000 [3], and WCDMA [4]) as well as Wi-Fi [5], the pseudo-random noise (PN) sequence [6]-[8] or its variants [4], [9]-[12] are used as the synchronization sequences. A complex PN sequence has a periodic autocorrelation of [4], [6], [10]

$$\gamma_{PN}(\kappa_1, \kappa_2) = \begin{cases} \frac{1}{N}\sum_{n=0}^{N-1} x[n+\kappa_1] x^*[n+\kappa_2] = -\frac{1}{N}, & \kappa_1 \neq \kappa_2 \\ 1, & \kappa_1 = \kappa_2 \end{cases}, \quad (1)$$

where $N$ is the period of the PN sequence. Therefore, the cyclically shifted PN sequences, i.e., sequences with different shift offsets (or lags) from the original PN sequence, have correlation of $-1/N$. PN sequences are thus commonly used as synchronization sequences, e.g., the downlink synchronization channel (SCH) and the uplink random access channel (RACH) in 2G cellular systems (IS-95) [1], [13] and 3G cellular systems (CDMA2000, WCDMA) [3], [4], [14], [15], and the preamble sequences in Wi-Fi.

Another class of sequences, the Zadoff-Chu (ZC) sequence, is a class of polyphase sequences defined as [16] [17] [18]

$$\mathbf{x}_\mu = \left\{ x_\mu(n) = e^{-j\frac{\pi \mu n(n+1)}{N}}, \ n = 0, 1, \cdots, N-1 \right\}, \quad (2)$$

where $N$ (odd) is the period of the sequence. The root of the sequence (or the sequence index) $\mu \in \{1, 2, \cdots, N-1\}$ is relatively prime to $N$.

The ZC sequence possesses an ideal or "perfect" periodic autocorrelation property (i.e., the periodic autocorrelation is zero for all shifts other than zero),

$$\gamma_{\mu\mu}(\Delta\kappa) = \frac{1}{N}\sum_{n=0}^{N-1} x_\mu[n] \cdot x_\mu^*[n+\Delta\kappa] = \delta[\Delta\kappa], \quad (3)$$

where the shift or lag $\Delta\kappa = -(N-1), \cdots 0, \cdots, N-1$. In (3) and in the rest of the paper, modulo-$N$ indexing is assumed.

Like PN sequences, ZC sequences typically have two primary applications in modern wireless communication systems. The first common application is in random multiple access where a set of cyclically-shifted ZC sequences (hence are orthogonal to each other) are used as random multiple access signals due to the orthogonality property between cyclically-shifted ZC sequence. In the LTE random access process, an access device randomly selects a ZC sequence from a given set of ZC sequences, and transmits it on the uplink random access resource to the access point. The access point detects the sequence using a bank of correlators, each of which is matched to a corresponding ZC sequence in the set. Unlike the PN sequences that suffer from the well-known near-far effect in multiple access, multiple devices (up to the size of the ZC sequence set) can transmit the sequences simultaneously on the same resource using a set of well-structured orthogonal ZC sequences [19]. Another application of ZC sequences can be found in time synchronization where a ZC sequence is employed as a time synchronization signal. In this case, the receiver detects the timing of the device by looking for the peak of the correlator output of the ZC sequence inside a timing hypothesis window. In LTE, ZC sequences are employed as the primary synchronization sequences (PSS) on the downlink to provide initial system time and frequency acquisition for an access device to synchronize its time and frequency to the system before it can start the random access process as described above. After the access point detects the random access sequence (i.e., a ZC sequence) on the uplink, it uses the detected timing of the access device from its access sequence to advance the device's transmit time to account for the round trip *propagation* delay such that uplink signals from different devices are aligned at the access point receiver. The cyclic prefix of the OFDM symbol thus only needs to accommodate the *multipath* delays from a device *after* the time alignment has been established [20]-[22]. Since the range of the propagation delay can be significantly larger than that of the mulitpath delay, significant overhead of cyclic prefix is saved and yet the interference to other uplink channels due to timing differences resulted from different propagation delays among devices can be avoided. This is a typical example that the ZC sequences are used as both random access signals and timing signals.

Although ZC sequences have been extensively studied, most of the earlier studies of ZC sequences assume no frequency offset between the transmitter and the receiver. The ideal periodic autocorrelation property of the ZC sequences is derived under this assumption [16], [18], [23]-[25]. Unfortunately, the frequency offset does have a profound impact on the autocorrelation property of a ZC sequence. Moreover, in practical wireless communication scenarios, the frequency offset between transceivers is inevitable due to the accumulated frequency uncertainties at the access device transmitter and the access point receiver as well as the Doppler spread resulting from the mobility of the access device [19], [26]. For example, during the random access process, the frequency of an access device is synchronized to the access point via the downlink SCH of the access point. The frequency offset on the uplink is thus determined by the access device frequency synthesizer error plus Doppler frequency. The frequency synthesizer error is typically within the range of 0.1 ppm or 200 Hz at a carrier frequency of 2 GHz [27]. In medium mobility scenarios, assuming an access device with a speed of 120 km/h at a carrier frequency of 2 GHz, the corresponding Doppler frequency is 222 Hz. Since the frequency synthesizer is synchronized to the downlink SCH, the Doppler effect is thus doubled on the uplink seen by the access point receiver, giving rise to a total of 644 Hz frequency offset at the access point, which is half of the OFDM symbol subcarrier spacing of LTE RACH (1.25 kHz).

Recently, the effect of the frequency offset on the autocorrelation of the ZC sequences is investigated [19] [21] [22] [28] [29] . In [19], and [28], the superiority of ZC sequences over PN sequences is clearly explained. The effect of frequency offset on the timing and detection performance of ZC sequences with different root indices is identified and analyzed. In [21], [22], and [28], ZC sequences are used for assisting OFDM symbol timing in the presence of a frequency offset where the timing uncertainty range is typically small (within a cyclic prefix range, e.g., $5\,\mu s$ ). In [29], the effect of frequency offset on the interference characteristics among multiple ZC sequences in an orthogonal ZC sequence set when used as multiple access sequences is studied, in which timing is assumed to be perfect. The aim of this paper is hence to study the effect of the frequency offset on the timing property of an individual ZC sequence, particularly the interplay between time and frequency uncertainties, when the ZC sequence is used as a general-purpose time synchronization sequence (where the time uncertain range can be large) using a new analytical framework. The proposed framework completely characterizes the timing behavior of any given ZC sequence, including the fundamental limitation (i.e., the irreducible timing error) which ultimately determines the maximum combined timing and frequency uncertainties that a ZC sequence can resolve as a synchronization sequence.

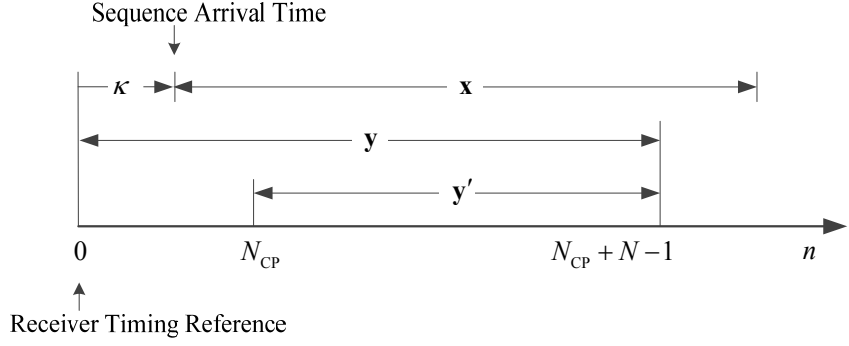

Fig. 1. Illustration of signal timing at a receiver, where $\kappa$ is the propagation delay from the transmitter to the receiver (or the arrival time relative to the receiver timing reference), and $N_{CP}$ denotes the cyclic prefix length of the OFDM symbol in samples.

Without loss of generality and for ease of discussion, in the sequel we continue to use the LTE RACH signal as a practical application model as in [29], but the focus is solely on the timing property of a ZC sequence as a time synchronization sequence under a frequency offset. The remainder of this paper is organized as follows. Section II analyzes the effect of the frequency offset on ZC sequence-based timing estimation. Section III provides more detailed analyses with numerical examples. Finally, the paper concludes in Section IV with a review of the main results.

## II. ANALYTICAL FRAMEWORK FOR TIMING ANALYSIS

In the section, we derive the analytical framework for analyzing the impact of a frequency offset on timing properties of a ZC sequence.

Assume sequence $\mathbf{x}$ is comprised of ZC sequence $\mathbf{s}$ of length $N$ with a cyclic prefix (CP) of length $N_{CP}$. An OFDM waveform that carries sequence $\mathbf{x}$ travels through the channel with gain $h$, and is received by a receiver with a frequency offset $\Delta f$ relative to the transmitter. The channel is assumed to be constant during the sequence transmission time. The sampled signal is denoted as $\mathbf{y}$ as shown in Fig. 1, whose sample element is represented as

$$y[n] = hx[n-\kappa]e^{j\frac{2\pi}{N}\Delta\lambda n} + \omega[n], \quad n = 0, 1, \cdots, N + N_{CP} - 1, \quad (4)$$

where $\Delta\lambda \triangleq \frac{\Delta f}{\Delta f_s}$, i.e., the frequency offset normalized to the OFDM subcarrier interval $\Delta f_s$, $\omega[n]$ is the independent complex Gaussian noise with zero mean and variance $\sigma^2$, i.e., $CN(0,\sigma^2)$. The arrival time of the sequence $\mathbf{x}$ at the receiver is $\kappa$, which corresponds to the actual timing of the transmitter seen at the receiver (cf. Fig. 1). Note that since at this point, the timing is not yet established between the transceivers, the cyclic prefix of the OFDM symbol waveform that carries the ZC sequence must be long enough to cover both propagation and multipath delays. The value of $N_{CP}$ is thus chosen such that $\max\{\kappa\} \leq N_{CP}$.

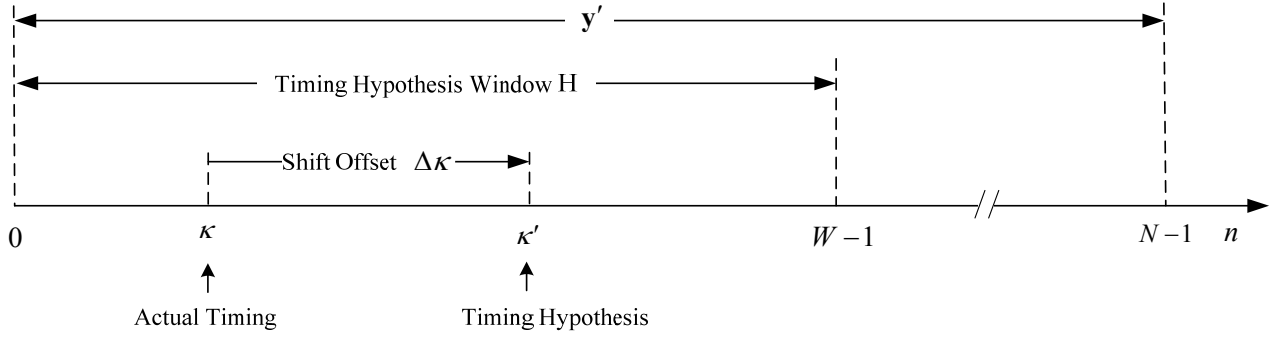

Fig. 2. Illustration of timing hypothesis window with respect to truncated (CP-removed) signal $y'[n]$, $n = 0, 1, \cdots, N-1$ $(W < N)$. Clearly, the timing hypothesis $\kappa'$ that corresponds to $\Delta\kappa = 0$, i.e., $\kappa' = \kappa$ is the correct timing hypothesis.

We now discard the first $N_{CP}$ samples and retain only the last $N$ samples from $\mathbf{y}$ to give $\mathbf{y}'$, namely,

$$\begin{aligned} y'[n] &\triangleq y[n + N_{CP}] \\ &= hx[n + N_{CP} - \kappa]e^{j\frac{2\pi}{N}\Delta\lambda(n+N_{CP})} + \omega[n + N_{CP}] \\ &= hs[n-\kappa]e^{j\frac{2\pi}{N}\Delta\lambda n} + w[n], \quad n = 0, 1, \cdots, N-1, \end{aligned} \tag{5}$$

where in the last step the constant phase $e^{j\frac{2\pi}{N}\Delta\lambda N_{CP}}$ is absorbed into $h$ (for notational simplification), and $w[n] \triangleq \omega[n + N_{CP}]$. The truncated signal $\mathbf{y}'$ contains a complete, but cyclically-shifted and noise-corrupted copy of ZC sequence $\mathbf{s}$. It is readily seen that the number of cyclic shifts is $\kappa$ that is unknown to the receiver.

To estimate the transmitter timing, i.e., to determine $\kappa$, the receiver performs $N$-point circular correlations of received signal $\mathbf{y}'$ with the original sequence $\mathbf{s}$ at multiple positions in a timing hypothesis window, $\{\kappa' \in H\}$, as illustrated in Fig. 2, where

$$H \triangleq \{0, 1, \cdots, W-1\} \tag{6}$$

is the timing hypothesis window, used to account for the time uncertainty, $\kappa$, and $W$ is the size of the hypothesis window, i.e., $W = |H|$, as depicted in Fig. 2. Clearly, the hypothesis window has to be large enough to cover the time uncertainty range, such that $\kappa \in H$. We will use timing hypothesis window and time uncertainty range (or window) interchangeably.

The output of the correlator at hypothesis $\kappa' \in H$ can be formulated as

$$\begin{aligned}
z(\kappa,\kappa',\Delta\lambda) &= \frac{1}{N}\sum_{n=0}^{N-1} y'[n+\kappa']\cdot s^*[n] \\
&= \left(\frac{1}{N}\sum_{n=0}^{N-1}\left(hs[n+\kappa'-\kappa]e^{j\frac{2\pi}{N}\Delta\lambda(n+\kappa')} + w[n+\kappa']\right)\cdot s^*[n]\right) \\
&= \gamma(\kappa'-\kappa,\Delta\lambda)he^{j\frac{2\pi}{N}\Delta\lambda\kappa'} + \upsilon_{\kappa'} \\
&= \gamma(\Delta\kappa,\Delta\lambda)he^{j\frac{2\pi}{N}\Delta\lambda\kappa'} + \upsilon_{\kappa'} \\
&\triangleq z_{\kappa'}(\Delta\kappa,\Delta\lambda),
\end{aligned} \quad (7)$$

where $\Delta\kappa \triangleq \kappa' - \kappa$ is the shift/time offset between the received ZC sequence $\mathbf{y}'$ and the original sequence $\mathbf{s}$ at timing hypothesis $\kappa'$ (cf. Fig. 2). Hence $\Delta\kappa = 0$ corresponds to $\kappa' = \kappa$, i.e., the correct timing hypothesis. In addition,

$$\gamma(\Delta\kappa,\Delta\lambda) \triangleq \frac{1}{N}\sum_{n=0}^{N-1} s[n+\Delta\kappa]s^*[n]e^{j\frac{2\pi}{N}\Delta\lambda n} \quad (8)$$

is the (circular) autocorrelation of ZC sequence $\mathbf{s}$ at shift offset or lag, $\Delta\kappa$, in the presence of frequency offset $\Delta\lambda$, and

$$\upsilon_{\kappa'} = \frac{1}{N}\sum_{n=0}^{N-1} s^*[n]w[n+\kappa']. \quad (9)$$

Clearly, $\upsilon_{\kappa'} \sim CN(0,\sigma^2/N)$. Since for $p \neq q$,

$$\begin{aligned}
E\{\upsilon_p\upsilon_q^*\} &= \frac{1}{N^2}E\left\{\sum_{n=0}^{N-1} s^*[n]w[n+p]\cdot \sum_{m=0}^{N-1} s[m]w^*[m+q]\right\} \\
&= \frac{1}{N^2}\sum_{n=0}^{N-1}\sum_{m=0}^{N-1} s^*[n]s[m]E\{w[n+p]w^*[m+q]\} \\
&= \frac{1}{N^2}\sum_{n=0}^{N-1}\sum_{m=0}^{N-1} s^*[n]s[m]\delta\big[(n+p)-(m+q)\big]\cdot\sigma^2 \\
&= \frac{\sigma^2}{N^2}\sum_{n=0}^{N-1} s^*[n]s[n+p-q] \\
&= 0.
\end{aligned} \quad (10)$$

Therefore, $\upsilon_{\kappa'}$ is identically and independently distributed (for different $\kappa' \in H$).

An estimate of timing, namely $\kappa$, is thus

$$\hat{\kappa} = \arg\max_{\kappa' \in H} |z_{\kappa'}(\Delta\kappa, \Delta\lambda)|^2. \tag{11}$$

The problem of timing detection for sequence $s$ can then be considered as a sequence correlation problem at the following shifts,

$$\Delta K_{\kappa}^{H} \triangleq \{-\kappa, \cdots, -1, 0, 1, \cdots, (W-1)-\kappa\}, \tag{12}$$

for a given $\kappa \in H$, where $\kappa$ is the actual sequence arrival time (cf. Fig. 2). (12) is thus a set that includes all possible shift offsets, $\Delta\kappa = \kappa' - \kappa$ corresponding to different timing hypothesis values of $\kappa'$ in H. The goal is thus to find the $\Delta\kappa \in \Delta K_{\kappa}^{H}$ that maximizes the timing detection metric $\zeta_{\kappa'}(\Delta\kappa, \Delta\lambda) \triangleq |z_{\kappa'}(\Delta\kappa, \Delta\lambda)|^2$.

The timing detection metric $\zeta_{\kappa'}(\Delta\kappa, \Delta\lambda)$ is independently (for different $\kappa' \in H$), non-central chi-square distributed, i.e.,

$$f_{\zeta_{\kappa'}(\Delta\kappa, \Delta\lambda)}(\zeta) = \frac{N}{\sigma^2} e^{-N\left(|\gamma(\Delta\kappa,\Delta\lambda)|^2 \frac{|h|^2}{\sigma^2} + \frac{\zeta}{\sigma^2}\right)} I_0\left(2N|\gamma(\Delta\kappa,\Delta\lambda)|\frac{|h|\sqrt{\zeta}}{\sigma^2}\right), \tag{13}$$

with degrees of freedom of two, a mean of

$$E\{\zeta_{\kappa'}(\Delta\kappa, \Delta\lambda)\} = |\gamma(\Delta\kappa, \Delta\lambda)|^2 |h|^2 + \frac{\sigma^2}{N}, \tag{14}$$

and a variance of

$$Var\{\zeta_{\kappa'}(\Delta\kappa, \Delta\lambda)\} = \frac{2\sigma^2}{N} |\gamma(\Delta\kappa, \Delta\lambda)|^2 |h|^2 + \frac{\sigma^4}{N^2}, \tag{15}$$

where

$$|\gamma(\Delta\kappa, \Delta\lambda)|^2 = \frac{1}{N^2} \sum_{n=0}^{N-1} \sum_{m=0}^{N-1} e^{-j\frac{2\pi}{N}(\mu\Delta\kappa - \Delta\lambda)(n-m)}$$
$$= |\text{sinc}(\Delta\lambda - \mu\Delta\kappa)|^2 \tag{16}$$

is the (squared) autocorrelation of ZC sequence $\mu$ at shift offset $\Delta\kappa$ and at frequency offset $\Delta\lambda$. Here the sinc function is defined as $\text{sinc}(x) = \frac{1}{N}\frac{\sin(\pi x)}{\sin(\pi x/N)}$. Clearly, in the absence of frequency offset, i.e., $\Delta\lambda = 0$, (16) is a perfect autocorrelation function that has a single non-zero value of one at $\Delta\kappa = 0$ and zero anywhere else, i.e., at $\Delta\kappa = \pm 1, \pm 2, \cdots, \pm(N-1)$.

It is seen that both the mean and variance are functions of the autocorrelation (16), which in turn is a function of a linear combination of $\Delta\lambda$ and $\Delta\kappa$, namely, $\Delta\lambda - \mu\Delta\kappa$.

We observe that the autocorrelation of a ZC sequence is a function of not only the shift offset but the frequency offset as well. The maximum correlation of one occurs when

$$\Delta\lambda = \mu\Delta\kappa + l \cdot N, \quad l \in \mathbb{Z}, \tag{17}$$

meaning that the maximum autocorrelation happens at lag $\Delta\kappa \neq 0$ and $\Delta\lambda \neq 0$. That is to say that the autocorrelation of a ZC sequence is no longer perfect in the presence of frequency offset. It implies that the detection metric, $\zeta_{\kappa'}(\Delta\kappa, \Delta\lambda)$ may become maximum at positions other than $\hat{\kappa} = \kappa$, thereby giving rise to false timing detections. In other words, the effective timing that the timing detector sees is a combination of both time and frequency offsets of the receiver relative to the ZC sequence signal. This unique property of the ZC sequence results in ambiguities between frequency offset and time offset, potentially degrading the perfect timing performance as promised by the perfect autocorrelation property of a ZC sequence.

In the following, we provide analytical derivations that give deeper insight into the interconnection between time and frequency uncertainties. We rewrite Equation (16) as

$$\left|\mathrm{sinc}(\Delta\lambda - \mu\Delta\kappa)\right|^2 = \left|\mathrm{sinc}(\Delta\lambda - \Delta\lambda^\dagger_{\Delta\kappa})\right|^2, \tag{18}$$

where

$$\Delta\lambda^\dagger_{\Delta\kappa} \triangleq \mu\Delta\kappa + l^\dagger N, \quad l^\dagger = \arg\min_{l \in \mathbb{Z}} \left|\mu\Delta\kappa + lN\right|. \tag{19}$$

We refer to $\Delta\lambda^\dagger_{\Delta\kappa}$ as the *critical frequency offset* of the autocorrelation function at a shift offset or lag, $\Delta\kappa \neq 0$, at which the autocorrelation has a maximum value of one. A critical frequency offset is thus the minimum frequency offset that shifts the maximum value of the autocorrelation from the zero shift offset to a non-zero location, $\Delta\kappa \neq 0$.

We define

$$\Lambda^\dagger_{\Delta\mathrm{K}} \triangleq \left\{\Delta\lambda^\dagger_{\Delta\kappa}, \Delta\kappa \in \Delta\mathrm{K}, \Delta\kappa \neq 0\right\}, \tag{20}$$

which includes all the critical frequency offsets of a ZC sequence for a given set of non-zero shift offsets of interest, $\Delta K$. We refer to these shift offsets as the *critical shift offsets*, and denote $\Delta \kappa^i$ as the critical shift offset corresponding to a critical frequency offset of value $i$.

We take the histogram of $\Lambda_{\Delta K}^{\dagger}$, to give

$$S(\Delta \lambda^{\dagger}), \quad \Delta \lambda^{\dagger} \in \Lambda_{\Delta K}^{\dagger}. \tag{21}$$

Hence, $S(\Delta \lambda^{\dagger})$ represents the multiplicity of $\Delta \lambda^{\dagger}$ in $\Lambda_{\Delta K}^{\dagger}$. When applied to timing detection with a hypothesis window $H$, we have

$$S(\Delta \lambda^{\dagger}), \quad \Delta \lambda^{\dagger} \in \Lambda_{\Delta K_{\kappa}^{H}}^{\dagger}, \tag{22}$$

where the shift offsets of interest are $\Delta K_{\kappa}^{H}$, defined in (12) for a given timing hypothesis window $H$ and $\forall \kappa \in H$.

We thus define

$$S_{H}(\Delta \lambda^{\dagger}) \triangleq \frac{1}{|H|} \sum_{\substack{\Delta \lambda_{\kappa}^{\dagger} = \Delta \lambda^{\dagger} \\ \Delta \lambda_{\kappa}^{\dagger} \in \left\{ \Lambda_{\Delta K_{\kappa}^{H}}^{\dagger}, \kappa \in H \right\}}} S(\Delta \lambda_{\kappa}^{\dagger}) \tag{23}$$

as the *timing spectrum* of a ZC sequence associated with a time uncertainty range H. It is clear that the timing spectrum of a ZC sequence is a function of the ZC sequence root parameter and the time uncertainty range.

It is intuitive that, for a good timing sequence, the critical frequency offsets $\Lambda_{\Delta K^{H}}^{\dagger} \triangleq \left\{ \Lambda_{\Delta K_{\kappa}^{H}}^{\dagger}, \kappa \in H \right\}$ associated with the shift offsets of a time uncertainty range $H$, namely,

$$\Delta K^{H} \triangleq \left\{ \Delta K_{\kappa}^{H}, \kappa \in H \right\} \tag{24}$$

must be much larger than the practical frequency error/uncertainty range of a given system. Therefore, it is natural to require that the timing spectrum of a ZC sequence contain as few spectral components at small critical frequency offsets as possible. We'll see more interconnections between the timing spectrum and timing properties of a ZC sequence in the next section.

From the above results, we can now obtain the timing detection error probability as a function of frequency offset $\Delta \lambda$. For a given $\kappa \in H$, the probability of $\Delta \kappa = \Delta \kappa^{*}$ is

$$p(\Delta\kappa = \Delta\kappa^*|\kappa)$$

$$= \int_0^\infty f_{\zeta_{\kappa'}(\Delta\kappa^*,\Delta\lambda)}(\zeta) \prod_{\substack{\Delta\kappa \neq \Delta\kappa^* \\ \Delta\kappa \in \Delta K_\kappa^H}} \left(1 - Q_1\left(\sqrt{2N}\left|\text{sinc}(\Delta\lambda - \mu\Delta\kappa)\right|\frac{|h|}{\sigma}, \frac{\sqrt{2N\zeta}}{\sigma}\right)\right) d\zeta$$

$$= \int_0^\infty \frac{N}{\sigma^2} e^{-N\left(\left|\text{sinc}(\Delta\lambda - \mu\Delta\kappa^*)\right|^2 \frac{|h|^2}{\sigma^2} + \frac{\zeta}{\sigma^2}\right)} I_0\left(2N\left|\text{sinc}(\Delta\lambda - \mu\Delta\kappa^*)\right|\frac{|h|}{\sigma}\sqrt{\frac{\zeta}{\sigma^2}}\right) \prod_{\substack{\Delta\kappa \neq \Delta\kappa^* \\ \Delta\kappa \in \Delta K_\kappa^H}} \left(1 - Q_1\left(\sqrt{2N}\left|\text{sinc}(\Delta\lambda - \mu\Delta\kappa)\right|\frac{|h|}{\sigma}, \frac{\sqrt{2N\zeta}}{\sigma}\right)\right) d\zeta$$

$$= \int_0^\infty N e^{-N\left(\left|\text{sinc}(\Delta\lambda - \mu\Delta\kappa^*)\right|^2 \eta + \frac{\zeta}{\sigma^2}\right)} I_0\left(2N\left|\text{sinc}(\Delta\lambda - \mu\Delta\kappa^*)\right|\sqrt{\eta}\sqrt{\frac{\zeta}{\sigma^2}}\right) \prod_{\substack{\Delta\kappa \neq \Delta\kappa^* \\ \Delta\kappa \in \Delta K_\kappa^H}} \left(1 - Q_1\left(\sqrt{2N}\left|\text{sinc}(\Delta\lambda - \mu\Delta\kappa)\right|\sqrt{\eta}, \sqrt{2N\frac{\zeta}{\sigma^2}}\right)\right) d\left(\frac{\zeta}{\sigma^2}\right)$$

$$= \int_0^\infty N e^{-N\left(\left|\text{sinc}(\Delta\lambda - \mu\Delta\kappa^*)\right|^2 \eta + \zeta\right)} I_0\left(2N\left|\text{sinc}(\Delta\lambda - \mu\Delta\kappa^*)\right|\sqrt{\eta\zeta}\right) \prod_{\substack{\Delta\kappa \neq \Delta\kappa^* \\ \Delta\kappa \in \Delta K_\kappa^H}} \left(1 - Q_1\left(\sqrt{2N}\left|\text{sinc}(\Delta\lambda - \mu\Delta\kappa)\right|\sqrt{\eta}, \sqrt{2N\zeta}\right)\right) d\zeta ,$$

(25)

where $\eta \triangleq \frac{|h|^2}{\sigma^2}$ is defined as the receive sample SNR, $Q_1$ is Marcum's $Q$ function. The total probability of $\Delta\kappa = \Delta\kappa^*$ is thus

$$p_{\Delta\kappa^*} = \sum_{\kappa = \max\{-\Delta\kappa^*, 0\}}^{\min\{W-1, W-\Delta\kappa^*-1\}} p(\kappa) \cdot p(\Delta\kappa = \Delta\kappa^*|\kappa)$$

$$= \frac{1}{W} \sum_{\kappa = \max\{-\Delta\kappa^*, 0\}}^{\min\{W-1, W-\Delta\kappa^*-1\}} p(\Delta\kappa = \Delta\kappa^*|\kappa),$$

(26)

where $p(\Delta\kappa = \Delta\kappa^*|\kappa)$ is defined in (25).

A timing error occurs when $\Delta\kappa = \Delta\kappa^* \neq 0$ is detected, which has the probability of

$$p_e = 1 - p_{\Delta\kappa^* = 0}$$

$$= 1 - \frac{1}{W} \sum_{\kappa=0}^{W-1} \int_0^\infty N e^{-N\left(\left|\text{sinc}(\Delta\lambda)\right|^2 \eta + \zeta\right)} I_0\left(2N\left|\text{sinc}(\Delta\lambda)\right|\sqrt{\eta\zeta}\right) \prod_{\substack{\Delta\kappa \neq 0 \\ \Delta\kappa \in \Delta K_\kappa^H}} \left(1 - Q_1\left(\sqrt{2N}\left|\text{sinc}(\Delta\lambda - \mu\Delta\kappa)\right|\sqrt{\eta}, \sqrt{2N\zeta}\right)\right) d\zeta.$$

(27)

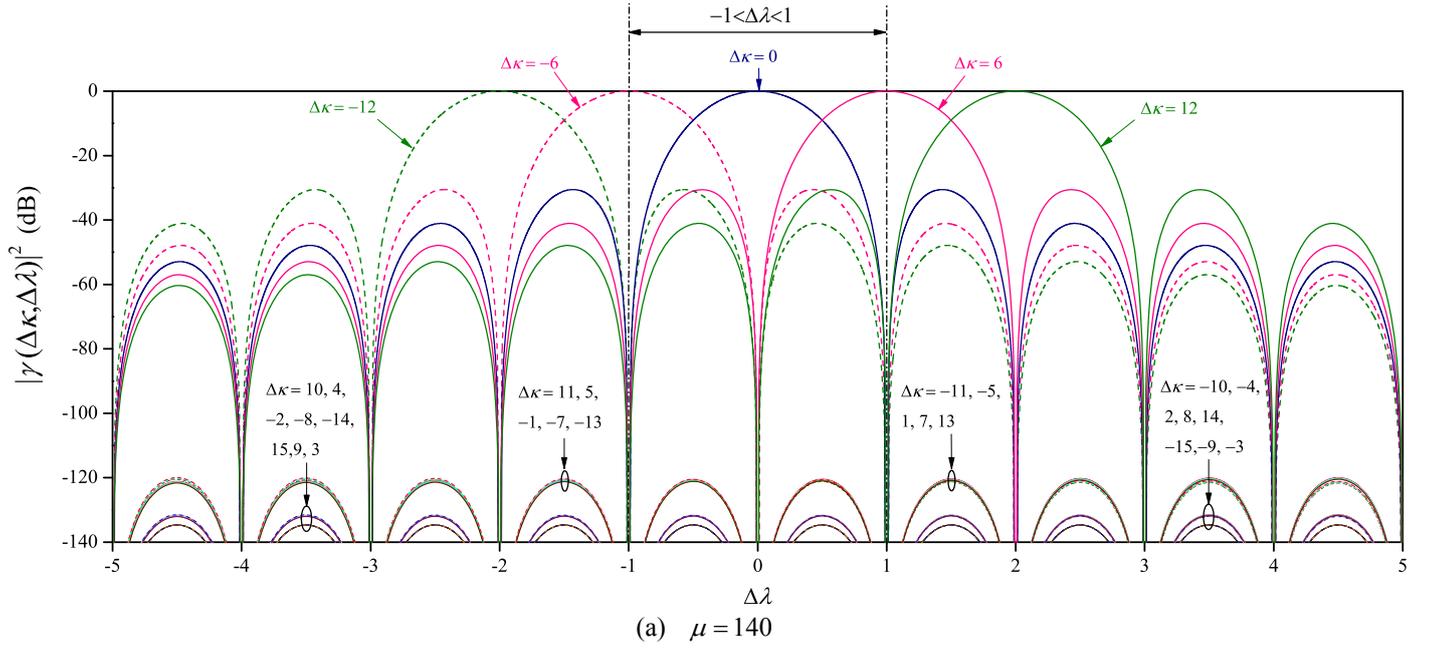

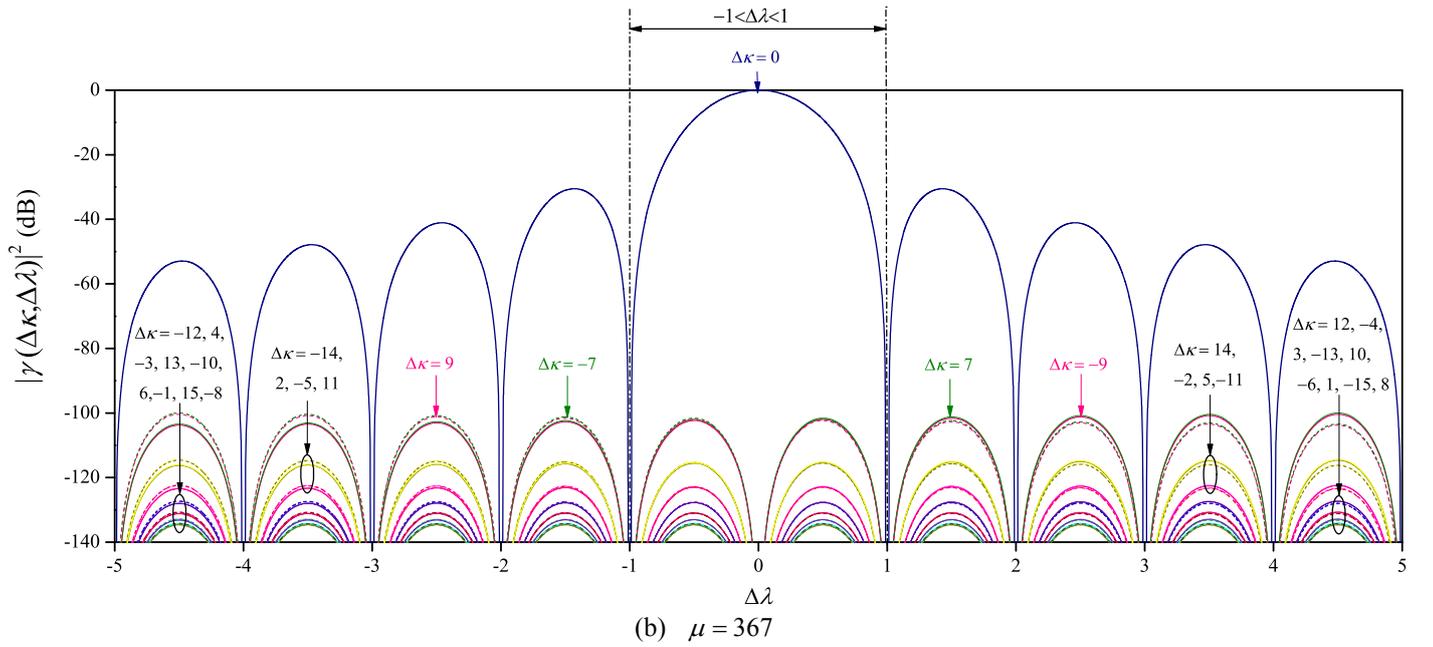

Fig. 3. Autocorrelation for ZC sequences $\mu = 140$ (a) and 367 (b) at the shift offsets $\Delta\kappa \in \{\Delta K_\kappa^H, \kappa \in U(0,15)\}$ as a function of the frequency offset. The solid curves correspond to the positive critical frequency offsets and the dotted curves the negative critical frequency offsets. TABLE I lists all the shift offsets for the time uncertainty window size of 16 and their associated critical frequency offsets.

TABLE I
THE SHIFT OFFSETS FOR THE TIME UNCERTAINTY WINDOW SIZE OF 16 AND THEIR ASSOCIATED CRITICAL FREQUENCY OFFSETS

| $\Delta\kappa$ | $\pm 1$ | $\pm 2$ | $\pm 3$ | $\pm 4$ | $\pm 5$ | $\pm 6$ | $\pm 7$ | $\pm 8$ | $\pm 9$ | $\pm 10$ | $\pm 11$ | $\pm 12$ | $\pm 13$ | $\pm 14$ | $\pm 15$ |
|---|---|---|---|---|---|---|---|---|---|---|---|---|---|---|---|
| $\Delta\lambda^\dagger$ ($\mu=140$) | $\pm 140$ | $\pm 280$ | $\mp 419$ | $\mp 279$ | $\mp 139$ | $\pm 1$ | $\pm 141$ | $\pm 281$ | $\mp 418$ | $\mp 278$ | $\mp 138$ | $\pm 2$ | $\pm 142$ | $\pm 282$ | $\mp 417$ |
| $\Delta\lambda^\dagger$ ($\mu=367$) | $\pm 367$ | $\mp 105$ | $\pm 262$ | $\mp 210$ | $\pm 157$ | $\mp 315$ | $\pm 52$ | $\pm 419$ | $\mp 53$ | $\pm 314$ | $\mp 158$ | $\pm 209$ | $\mp 263$ | $\pm 104$ | $\mp 368$ |

III. NUMERICAL EXAMPLES AND ANALYSIS

In this section, we use detailed examples to help better understand and validate the analytical results developed in the previous section. We use two *exemplary* ZC sequences (one with $\mu = 140$ and one $367$, both of length $N = 839$ — the same length used in LTE RACH) to show how their distinct timing performance is related to their unique timing spectral properties.

First we show in Fig. 3 the autocorrelation function of a ZC sequence as a function of the frequency offset $\Delta\lambda$ at the specific shift offsets associated with a timing hypothesis window size of $|\mathrm{H}| = W = 16$. This timing hypothesis window size provides a tolerance to a time uncertainty up to 16 samples or ~10 μs. For the random access application, this uncertainty also translates to a cell size of ~1000m [19]. If we assume the frequency error or frequency uncertainty in a typical wireless application is less than 1.25 kHz, i.e., in the range of $|\Delta f| < 1.25$ kHz or $|\Delta\lambda| < 1$ under the LTE RACH framework, $|\mathrm{sinc}(\Delta\lambda - \mu\Delta\kappa)|^2 \neq 0$ for $|\Delta\lambda| \neq 0$, due to the fact that the side ripples of the *sinc* function in (16) cause energy leakage between lags. For every $\Delta\kappa \in \Delta\mathrm{K}$ ($\Delta\kappa \neq 0$), there is a corresponding critical frequency offset $\Delta\lambda^\dagger$ at which a maximum correlation coefficient of one occurs.

In Fig. 3 (a), i.e., $\mu = 140$, $|\Delta\kappa^{\pm 1}| = 6$ corresponds to a critical frequency offset of $|\Delta\lambda^\dagger| = 1$ which is the smallest critical frequency offset among all possible critical frequency offsets. At critical shift offset $|\Delta\kappa^{\pm 1}| = 6$, any small frequency offset incurs a large autocorrelation value. This value even becomes greater than that at $\Delta\kappa = 0$ when frequency offset $\Delta\lambda > 0.5$ for $\Delta\kappa^1 = 6$ or when $\Delta\lambda < -0.5$ for $\Delta\kappa^{-1} = -6$. The autocorrelation can also be non-zero at other non-zero shift offsets in the frequency offset range of $|\Delta\lambda| < 1$ ($|\Delta\lambda| \neq 0$), e.g., at the critical shift offsets $|\Delta\kappa^{\pm 2}| = 12$ (corresponding to a critical frequency offset of $|\Delta\lambda^\dagger| = 2$) and at $|\Delta\kappa^{\pm 138}| = 11$ (corresponding to $|\Delta\lambda^\dagger| = 138$), but with smaller order of magnitudes. It is seen that the non-zero shift offset, $\Delta\kappa \neq 0$, associated with a smaller critical frequency offset (e.g., $|\Delta\lambda^\dagger| = 1$) has a larger effect on autocorrelation than that with a greater critical frequency offset (e.g., $|\Delta\lambda^\dagger| = 2$ or

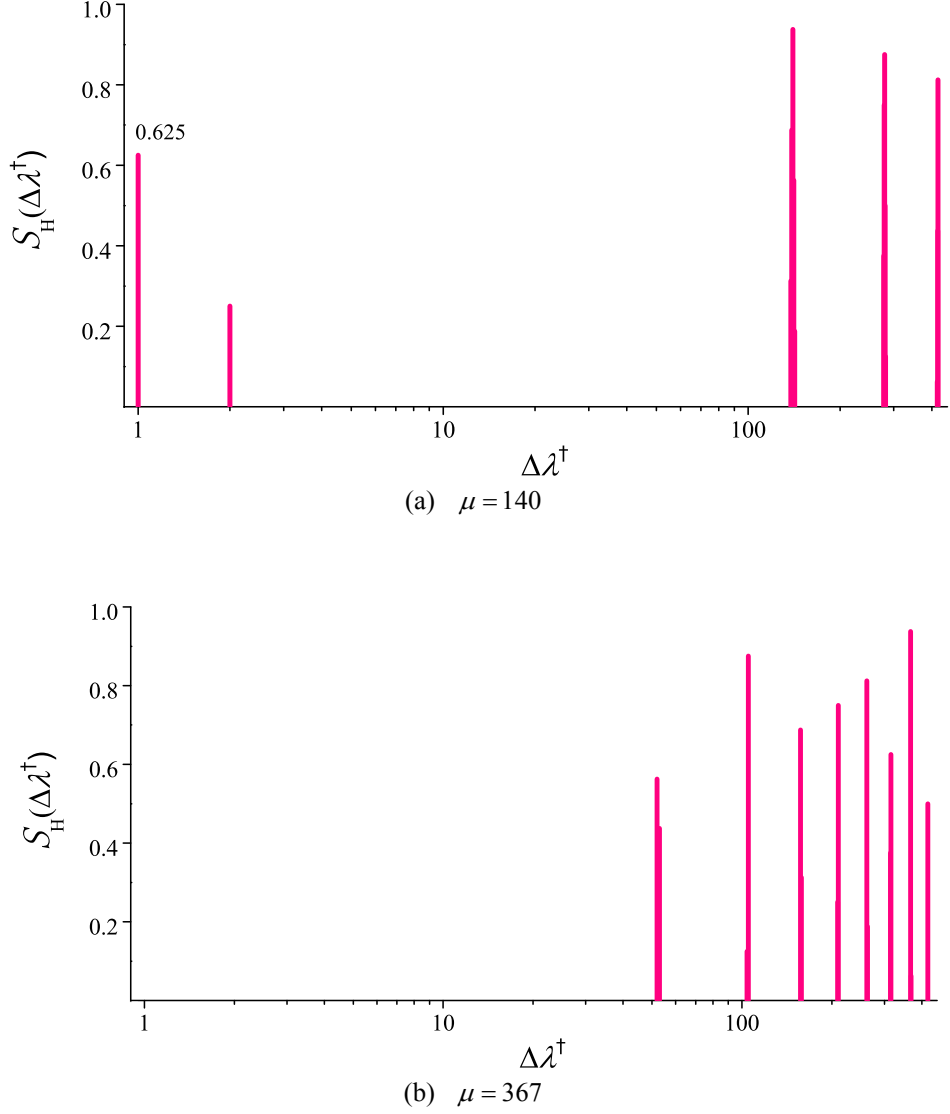

Fig. 4. The timing spectra for ZC sequences $\mu = 140$ (a) and 367 (b), and time uncertainty range of $|H| = 16$. Only the positive half ($\Delta\lambda^\dagger > 0$) of the spectrum is plotted since the spectrum is symmetric.

$|\Delta\lambda^\dagger| = 138$). Therefore, mistiming is more likely to happen at the non-zero shift offsets with smaller magnitude of critical frequency offsets.

Whereas for the ZC sequence with $\mu = 367$ as shown in Fig. 3 (b) and TABLE I, the smallest critical frequency offset associated with the non-zero shift offsets is $|\Delta\lambda^\dagger| = 52$, whose autocorrelation value is below -100 dB within $|\Delta\lambda| < 1$. Therefore, $|\Delta\kappa^{\pm 52}| = 7$ and other non-zero shift offsets have little effect on the timing performance.

This important point can be better represented by the timing spectrum. The timing spectra for $\mu = 140$

and 367 with timing hypothesis window size $|\mathrm{H}|=W=16$ are plotted in Fig. 4. For $\mu=140$, we see four spectral components (two at the negative side of the spectrum) at small critical frequency offsets of $\Delta\lambda^\dagger=\pm 1$ and $\pm 2$ with respect to $\Delta\kappa=\pm 6$ and $\pm 12$. We thus expect high mistiming probabilities at the above four offsets. Whereas for $\mu=367$, no spectral component at the lower end of the critical frequency offset is observed. Better timing performance from ZC sequence $\mu=367$ can thus be expected.

The above phenomena can be further verified by examining the mean of the detection metric at offset $\Delta\kappa\neq 0$ (the offset corresponding to the mistiming, cf. Fig. 2) relative to the mean of the detection metric at offset $\Delta\kappa=0$ (the shift offset corresponding to the correct timing) in the timing hypothesis window against frequency offset $\Delta\lambda$. From (14), this relative mean detection metric is given by

$$\frac{E\{\zeta_{\kappa'}(\Delta\kappa,\Delta\lambda)|\Delta\kappa\neq 0\}}{E\{\zeta_{\kappa'}(\Delta\kappa,\Delta\lambda)|\Delta\kappa=0\}} = \frac{|\mathrm{sinc}(\Delta\lambda-\mu\Delta\kappa)|^2|h|^2+\frac{\sigma^2}{N}}{|\mathrm{sinc}(\Delta\lambda)|^2|h|^2+\frac{\sigma^2}{N}}$$

$$= \frac{N|\mathrm{sinc}(\Delta\lambda-\mu\Delta\kappa)|^2+\eta^{-1}}{N|\mathrm{sinc}(\Delta\lambda)|^2+\eta^{-1}}. \tag{28}$$

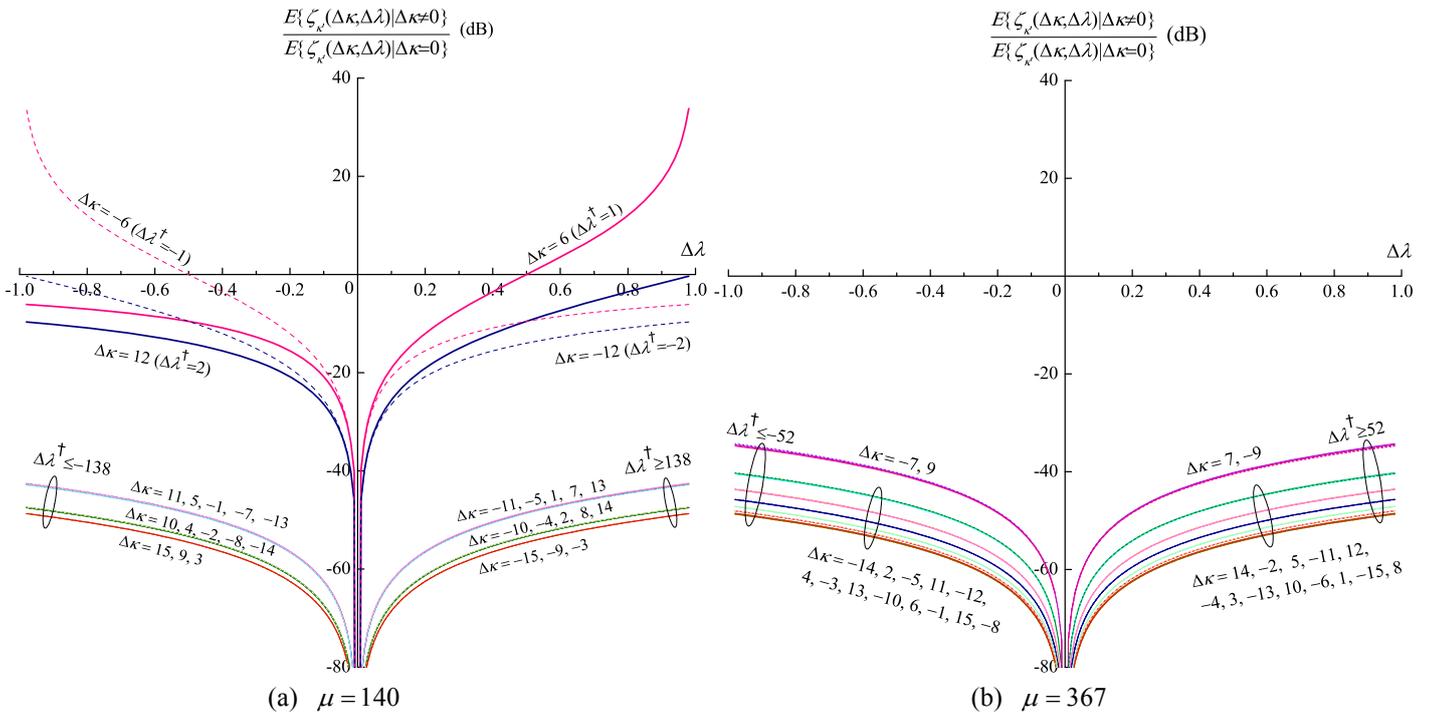

Fig. 5. The mean detection metric value (at high SNR) at non-zero shift offset $\Delta\kappa\neq 0$ (error) relative to $\Delta\kappa=0$ (correct) against frequency offset $\Delta\lambda$ for ZC sequences $\mu=140$ (a) and 367 (b). The solid curves correspond to the positive critical frequency offsets and the dotted curves the negative critical frequency offsets. The timing window size is $|\mathrm{H}|=16$.

At high SNR, i.e., $\eta = \frac{|h|^2}{\sigma^2} \to \infty$, this ratio becomes

$$\frac{E\{\zeta_{\kappa'}(\Delta\kappa, \Delta\lambda)|\Delta\kappa \neq 0\}}{E\{\zeta_{\kappa'}(\Delta\kappa, \Delta\lambda)|\Delta\kappa = 0\}} \to \frac{|\text{sinc}(\Delta\lambda - \mu\Delta\kappa)|^2}{|\text{sinc}(\Delta\lambda)|^2}. \tag{29}$$

The larger this ratio is, the more likely it is for the wrong timing offset (i.e., $\Delta\kappa \neq 0$) to be detected. In particular, when this ratio becomes greater than one (0 dB), the detection energy for the mistiming offset, $E\{\zeta_{\kappa'}(\Delta\kappa, \Delta\lambda)|\Delta\kappa \neq 0\}$, is greater than that of the actual timing, $E\{\zeta_{\kappa'}(\Delta\kappa, \Delta\lambda)|\Delta\kappa = 0\}$, giving rise to a non-zero error rate regardless of how high the SNR is (i.e., an *irreducible* detection error). Next, we use the two ZC sequence examples to further elaborate on this fundamental limitation of a ZC sequence.

Fig. 5 (a) plots (29) for $\mu = 140$ and $W = 16$. It is observed that for $\Delta\kappa = 6$ (corresponding to a critical frequency offset $\Delta\lambda^\dagger = 1$), the ratio $\frac{E\{\zeta_{\kappa'}(\Delta\kappa, \Delta\lambda)|\Delta\kappa = 6\}}{E\{\zeta_{\kappa'}(\Delta\kappa, \Delta\lambda)|\Delta\kappa = 0\}}$ (solid red line) increases as $\Delta\lambda$ ($\Delta\lambda > 0$) increases and becomes larger than zero (dB) at $\Delta\lambda > 0.5$, agreeing well with the observation from Fig. 3. That is, the mean energy of the correlator output at $\Delta\kappa = 6$ is greater than that at $\Delta\kappa = 0$, i.e., $E\{\zeta_{\kappa'}(\Delta\kappa, \Delta\lambda)|\Delta\kappa = 6\} > E\{\zeta_{\kappa'}(\Delta\kappa, \Delta\lambda)|\Delta\kappa = 0\}$ for $\Delta\lambda > 0.5$, thereby producing irreducible timing errors. For the case of $\Delta\kappa = 12$ (corresponding to $\Delta\lambda^\dagger = 2$), it is seen that $\frac{E\{\zeta_{\kappa'}(\Delta\kappa, \Delta\lambda)|\Delta\kappa = 12\}}{E\{\zeta_{\kappa'}(\Delta\kappa, \Delta\lambda)|\Delta\kappa = 0\}} < \frac{E\{\zeta_{\kappa'}(\Delta\kappa, \Delta\lambda)|\Delta\kappa = 6\}}{E\{\zeta_{\kappa'}(\Delta\kappa, \Delta\lambda)|\Delta\kappa = 0\}}$, $\forall |\Delta\lambda| < 1$, since the critical frequency offset $|\Delta\lambda^\dagger|$ corresponding to $\Delta\kappa = 12$ is greater than that for $\Delta\kappa = 6$, hence less effect on timing as compared to $\Delta\kappa = 6$, as will be seen later. For other shift offsets whose corresponding timing offsets are within the hypothesis window, $\Delta\kappa \in H$ ($\Delta\kappa > 0$), the corresponding ratios are much smaller than 0 dB since the magnitude of the corresponding critical frequency offsets is much larger than one. Thereby, their effect on timing is much less significant. Similar observations can be made for $\Delta\kappa < 0$.

It hence becomes clear that the actual value of the residual error probability when $\Delta\lambda > 0.5$ simply depends on the probability that $\Delta\kappa^I \in \Delta K^H$ occurs in the hypothesis window. Referring to (24) and (12), and assuming the arrival time is uniformly distributed in H, the error probability has a floor at

$$\tilde{p}_e = \frac{|H| - |\Delta\kappa^{\pm 1}|}{|H|} = \frac{W - |\Delta\kappa^{\pm 1}|}{W}, \tag{30}$$

when $\Delta\lambda > 0.5$ due to $\Delta\kappa^1 \in \Delta K^H$ or $\Delta\lambda < -0.5$ due to $\Delta\kappa^{-1} \in \Delta K^H$.

It can be shown that $\tilde{p}_e$ is nothing but the magnitude of the timing spectral component at the critical frequency offsets of $|\Delta\lambda^\dagger| = 1$. That is, the timing spectrum of a ZC sequence indicates the actual timing error floor of the corresponding ZC sequence. This useful property of the timing spectrum determines the ultimate limitation of a ZC sequence as a timing sequence.

As a marginal case when $\Delta\lambda = 0.5$, the spurious peak at $\Delta\kappa^1$ shares the same strength as the one at $\Delta\kappa = 0$, the error floor therefore halves that for $\Delta\lambda > 0.5$, i.e.,

$$\tilde{p}_e = \frac{|H| - |\Delta\kappa^{\pm 1}|}{2|H|} = \frac{W - |\Delta\kappa^{\pm 1}|}{2W}, \tag{31}$$

at $\Delta\lambda = \pm 0.5$ if $\Delta\kappa^{\pm 1} \in \Delta K^H$. This case is only of theoretical interest though since the probability of a frequency offset occurs at the exact value of 0.5 is zero.

For the example of $\mu = 140$ and $W = 16$, we have $|\Delta\kappa^{\pm 1}| = 6$. The predicted error floor by (30) is $\tilde{p}_e = 0.625$ for all $|\Delta\lambda| > 0.5$, which is the same as by the timing spectrum in Fig. 4 (a).

Whereas for $\mu = 367$ shown in Fig. 5 (b), we observe that $\dfrac{E\{\zeta_{\kappa'}(\Delta\kappa, \Delta\lambda) | \Delta\kappa \neq 0\}}{E\{\zeta_{\kappa'}(\Delta\kappa, \Delta\lambda) | \Delta\kappa = 0\}} \ll 0$ (dB), as a result of lack of spectral components in the lower frequency offset region as shown in Fig. 4 (b). Furthermore, since the spectral component of $\mu = 367$ at $\Delta\lambda^\dagger = 1$ is zero, i.e., $\Delta\kappa^1 \notin \Delta K^H$, the frequency offset in this case is not expected to introduce a timing error floor. ZC sequence $\mu = 367$ is hence a better choice than $\mu = 140$ for timing applications with the time uncertainty range of 16 (samples).

In the following, we validate the above analytical results via simulations.

Fig. 6 plots the timing probability distribution at the shift offset of $\Delta\kappa \in \Delta K^H$ calculated from (26) at the frequency offset of $\Delta\lambda = 0.5$. With $\mu = 140$, there are four spurious peaks in the timing hypothesis window in addition to the peak at $\Delta\kappa = 0$ (i.e., the correct shift offset). The four spurious ones are at $\Delta\kappa = \pm 6$ and $\pm 12$, corresponding to the four (two at the two positive critical frequency offsets and two at the negative ones) lower end spectral components in Fig. 4 (a). More importantly, just as the timing spectral component at $\Delta\lambda^\dagger = 1$ indicates an *irreducible* timing error rate of 0.625 for any frequency offset $\Delta\lambda > 0.5$, Fig. 7 (a) shows the detection error rate plateaus at 0.625 at $\Delta\lambda = 0.6$ and 0.7, and floors at half of that at $\Delta\lambda = 0.5$. The PN sequence on the other hand only has one dominant peak at $\Delta\kappa = 0$, and hence outperforms the ZC sequence in this case. As for the ZC sequence of $\mu = 367$, there is only one dominant peak present at $\Delta\kappa = 0$, as a consequence of lack of lower end spectral components present in its timing spectrum in Fig. 4 (b), resulting in a timing performance similar to PN (cf. Fig. 7 (b)).

However, it does not necessarily mean that ZC sequence $\mu = 367$ performs as well in even wider time uncertainty ranges. It can be seen from the fact that the timing spectrum is not only a function of the root parameter of the ZC sequence but also a function of the time uncertainty at the receiver. To demonstrate this point, we increase the timing hypothesis window size of ZC sequence $\mu = 367$ from $|H| = 16$ to 20, corresponding to the time uncertainty window of 20 samples or ~13 μs. The resultant timing spectrum is plotted in Fig. 8. We now see a new spectral component with a magnitude of 0.2 appear at the critical frequency offset of $\Delta\lambda^\dagger = 1$ (and $\Delta\lambda^\dagger = -1$ by symmetry) corresponding to shift offsets of $\Delta\kappa = -16$ and 16, respectively, i.e., $\Delta\kappa^{\pm 1} = \mp 16 \in \Delta K^H$. High mistiming probabilities at the above two shift offsets, particularly an error floor of 0.2 for $|\Delta\lambda| > 0.5$, are therefore expected, which is further verified by (27) and simulations in Fig. 9, where two spurious detection peaks occur at exactly these two locations, causing severely deteriorated timing detection performance with an error floor at 0.2 for $\Delta\lambda > 0.5$ (and 0.2/2=0.1 for $\Delta\lambda = 0.5$) as shown in Fig. 10. We thus see that the same ZC sequence can have very different timing spectra, and hence very different timing performance, in different time uncertainty ranges.

Fig. 11-Fig. 13 show the timing performance of another ZC sequence with $\mu = 29$ for a timing uncertainty window size $|H| = 20$. In Fig. 11, the lack of spectral components in the lower end indicates a good timing performance. In particular, the absence of a spectral component at $\lambda^\dagger = \pm 1$ guarantees no timing error floor within the given time uncertainty window, which is further confirmed by the results in Fig. 12 and Fig. 13.

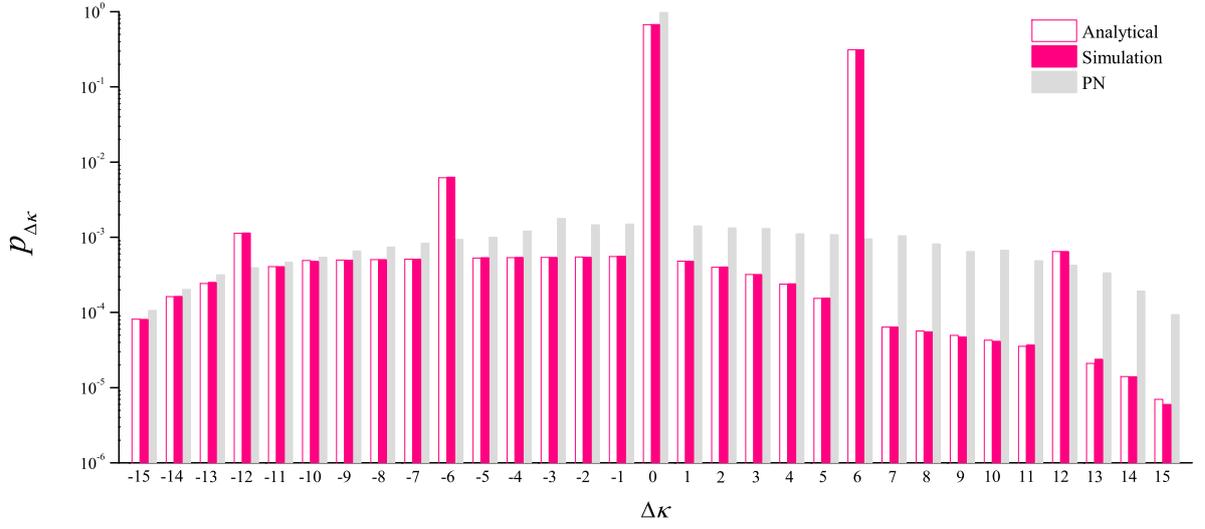

(a) $\mu = 140$

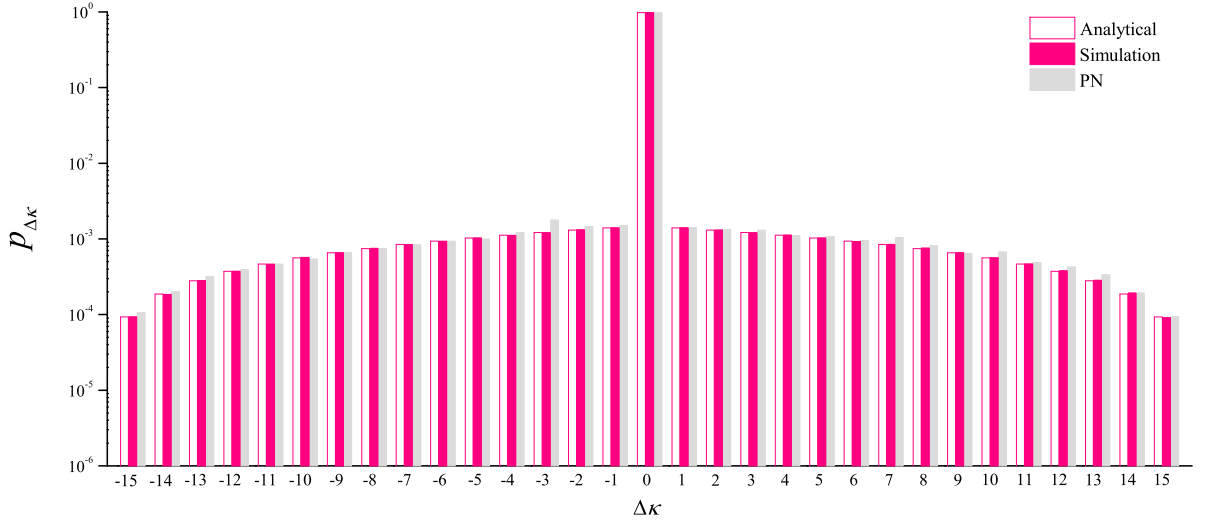

(b) $\mu = 367$

Fig. 6. Timing detection probability distribution, i.e., the detection error probability at shift offset $\Delta\kappa \in \Delta K_\kappa^H$ for ZC sequences $\mu$=140 (a) and 367 (b). In the simulation, the sequence arrival time $\kappa$ is a random variable, and uniformly distributed over time uncertainty window H, i.e., $\kappa \in U(0, W-1)$, where the window size is $|H| = W = 16$. The results are collected over various values of $\kappa$. The frequency offset is $\Delta\lambda = 0.5$, and receive sample SNR is $\eta = -15$ dB. The simulation results are plotted against the analytical result in (26). The results from PN sequences are also shown. The PN sequence is truncated to match the ZC sequence length, i.e., $N = 839$. The natural period of the PN sequence is $2^{25} - 1$.

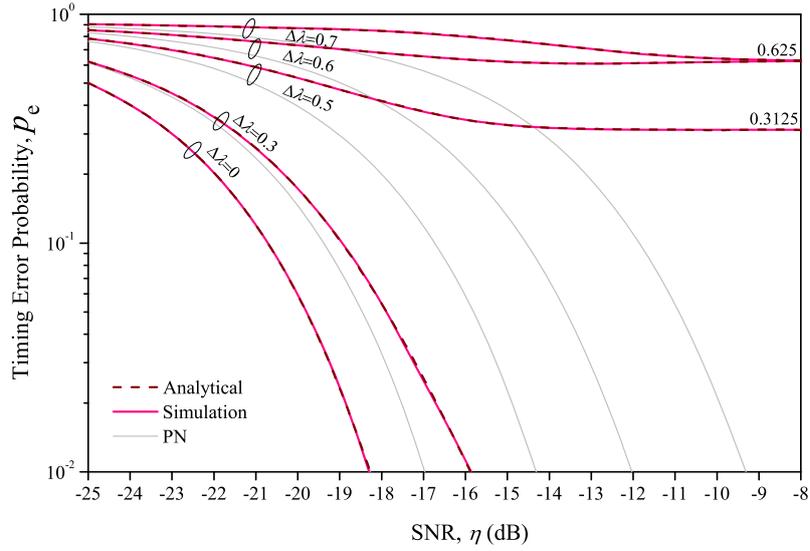

(a)  $\mu = 140$

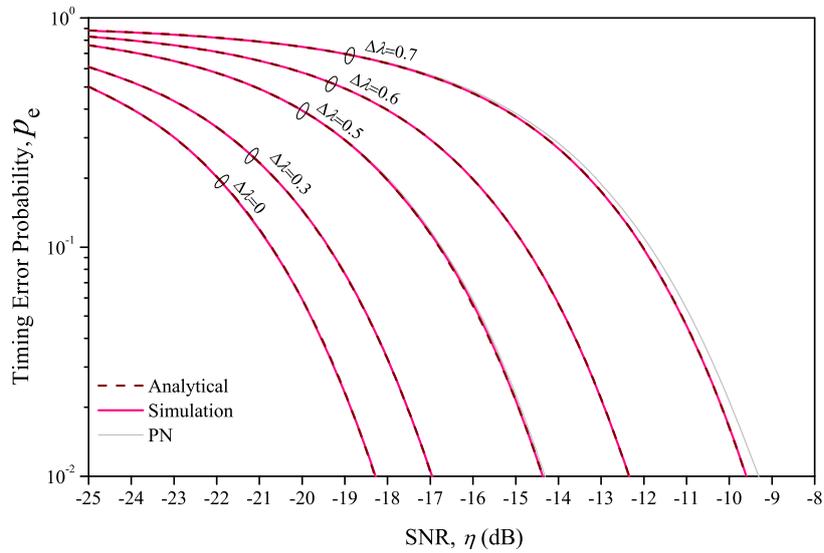

(b)  $\mu = 367$

Fig. 7. Timing detection performance (timing error probability) of ZC sequences  $\mu = 140$  (a) and 367 (b) at various frequency offsets. The timing hypothesis window size is 16. The timing error probabilities from the simulation are plotted against the analytical result given in (27). It is observed that there is a timing error floor at 0.625 as the frequency offset $\Delta\lambda$ becomes greater than 0.5 for  $\mu = 140$  and no floor for 367, just as indicated by the timing spectra in Fig. 4 (a) and Fig. 4 (b), respectively. The PN simulation results (gray) are also plotted for reference.

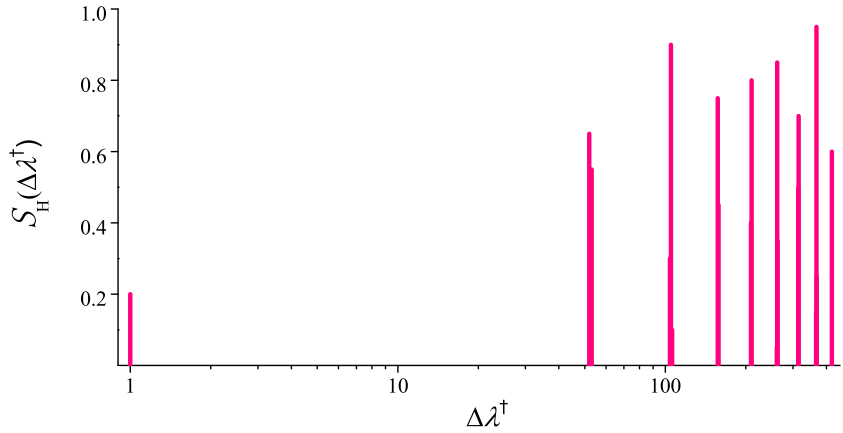

Fig. 8. The timing spectrum for ZC sequence $\mu = 367$ with timing hypothesis window size of $|\mathrm{H}| = W = 20$. Only the positive half ($\Delta\lambda^\dagger > 0$) of the spectrum is plotted due to the symmetry of the spectrum.

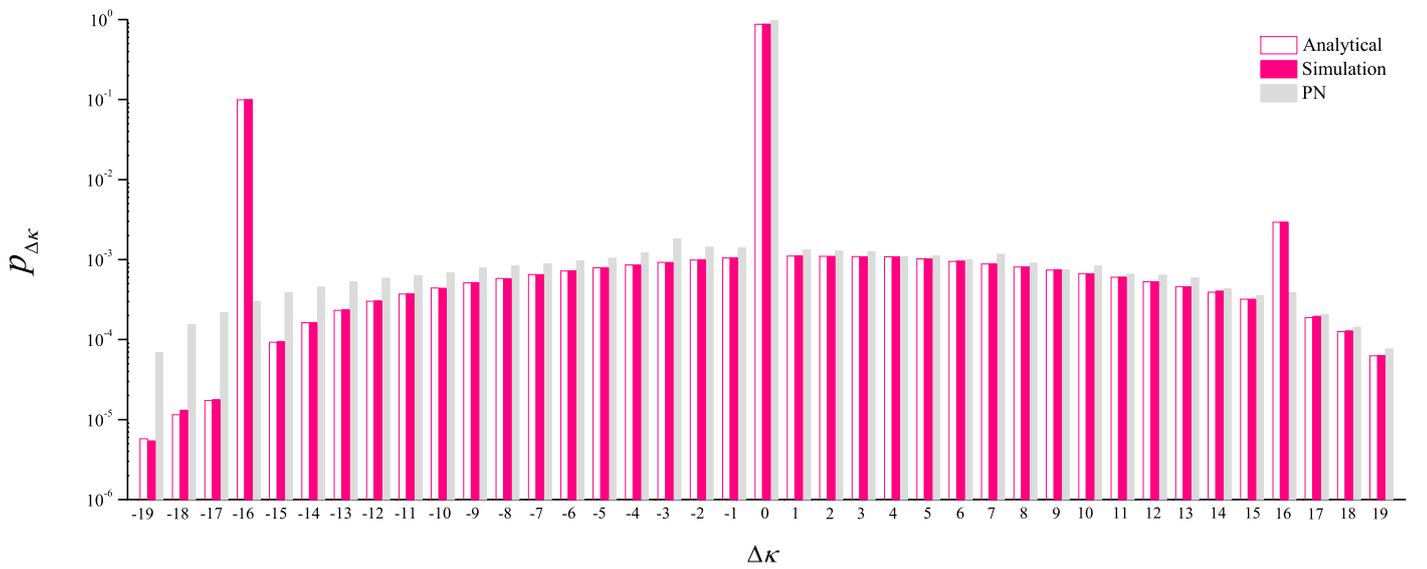

Fig. 9. Timing detection probability distribution (shown as shift offset $\Delta\kappa$) over $\kappa \in U(0, W-1)$ for ZC sequence $\mu = 367$ with timing hypothesis window size of $|\mathrm{H}| = W = 20$. As in Fig. 6, the frequency offset is $\Delta\lambda = 0.5$; the receive sample SNR of $-15$ dB; and the PN sequence is truncated to match the ZC sequence length, i.e., $N = 839$. The natural period of the PN sequence is $2^{25} - 1$.

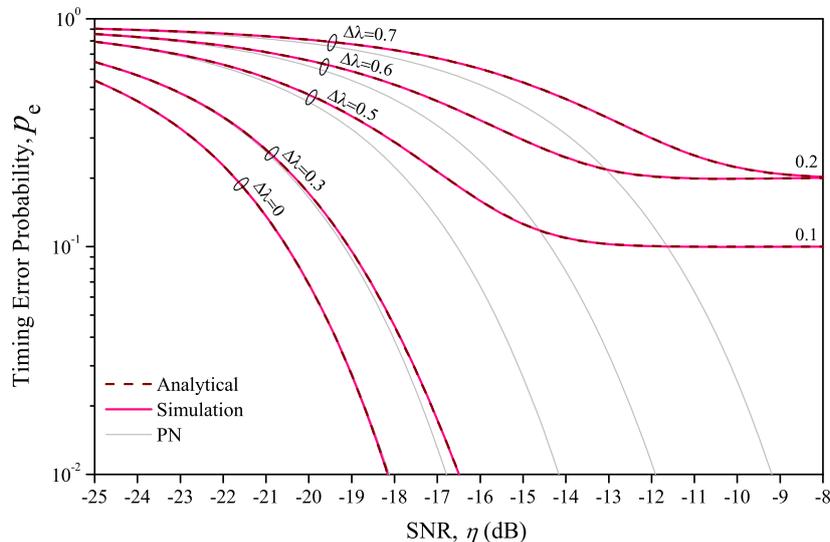

Fig. 10. Timing detection performance (timing error probability) of a ZC sequence ($\mu = 367$) at various frequency offsets. The timing hypothesis window size is 20. There is an error floor at 0.2 as the frequency offset $\Delta\lambda$ becomes greater than 0.5, the same value as indicated by the timing spectrum in Fig. 8. The PN simulation results (gray) are also plotted for reference.

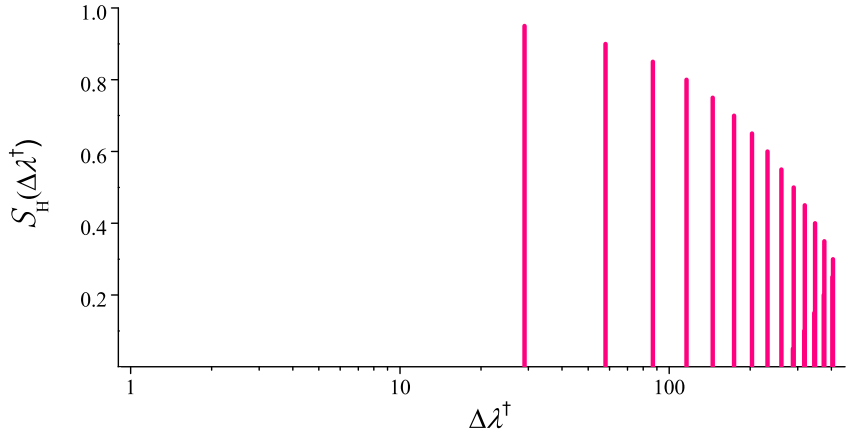

Fig. 11. The timing spectrum for ZC sequence $\mu = 29$ with timing hypothesis window size of $|H| = W = 20$. Only the positive half ($\Delta\lambda^\dagger > 0$) of the spectrum is plotted due to the symmetry of the spectrum.

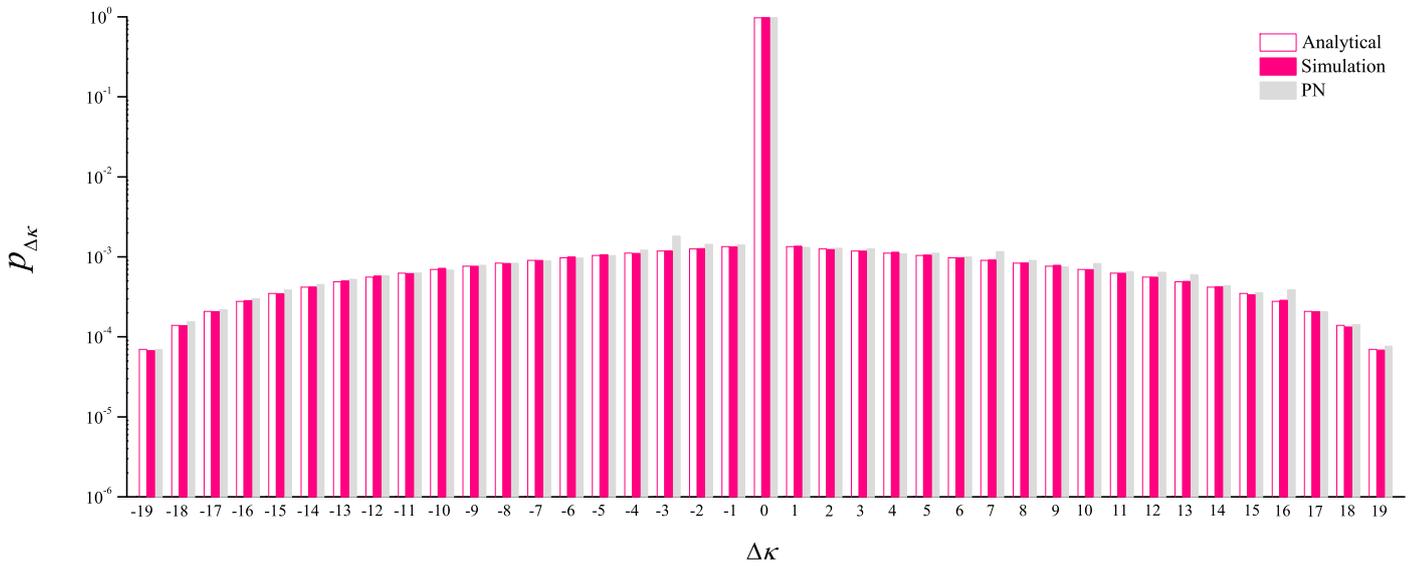

Fig. 12. Timing detection probability distribution (shown as shift offset $\Delta\kappa$) over $\kappa \in U(0, W-1)$ for ZC sequence $\mu = 29$ with timing hypothesis window size of $|H| = W = 20$. The frequency offset is $\Delta\lambda = 0.5$; the receive sample SNR of -15dB; and the PN sequence is truncated to match the ZC sequence length, i.e., $N = 839$. The natural period of the PN sequence is $2^{25} - 1$.

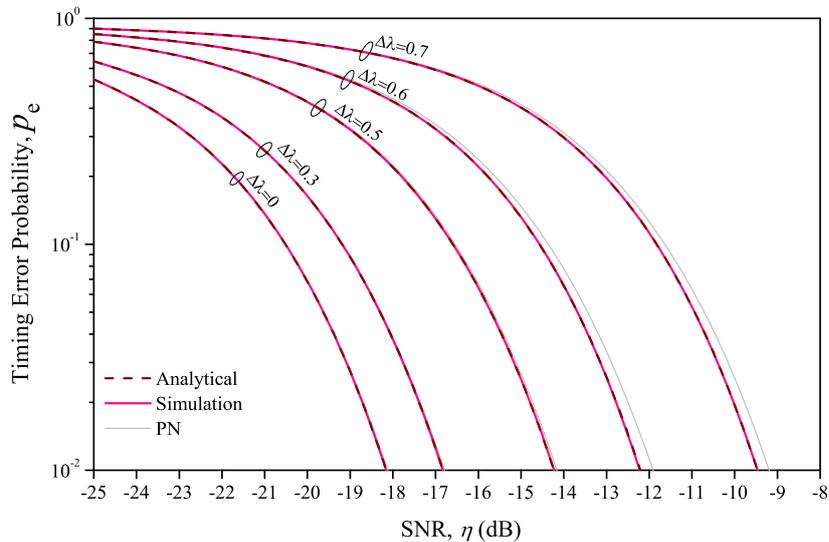

Fig. 13. Timing detection performance (timing error probability) of a ZC sequence ($\mu = 29$) at various frequency offsets. The timing hypothesis window size is 20. There is no timing error floor, which agrees with the timing spectrum in Fig. 11. The PN simulation results (gray) are also plotted for reference.

IV. CONCLUSION

A ZC sequence is well known for its perfect autocorrelation properties. As a result, ZC sequences have found their wide applications in modern cellular systems as synchronization sequences, such as the LTE Primary Synchronization Signal (PSS) and the Random Access Channel (RACH) signal. However, the perfect time autocorrelation is in general not true in practical applications when a frequency synchronization error is present between the transmitter and the receiver. That is, the perfect autocorrelation property is lost under non-zero frequency offset, consequently degrading the timing performance of a ZC sequence. The severity of the degradation depends on both the root parameter of a ZC sequence and the time uncertainty. It thus signifies the necessity for understanding the behavior and performance of a ZC sequence when applied to practical communication systems. In this paper, we develop an analytical framework that characterizes the timing properties of a ZC sequence. In particular, we introduce a concept of timing spectrum that relates the frequency offset effect on a ZC sequence timing performance directly to the root parameter of a ZC sequence combined with the time uncertainty at a receiver. We demonstrate how the time uncertainty and the selection of the root parameter can jointly affect the sensitivity of a ZC sequence's timing performance to a frequency offset and how a timing spectrum fully characterizes this property including the irreducible timing error probability, which ultimately determines the maximum combined timing and frequency uncertainties that a ZC sequence can resolve as a synchronization sequence. The analytical framework proposed in this paper provides an effective analytical tool for timing signal design and performance analysis.

REFERENCES

[1] F. Khan, *LTE for 4G Mobile Broadband: Air Interface Technologies and Performance*, Cambridge University, 2009.

[2] *Mobile Station–Base Station Compatibility Standard for Dual-Mode Wideband Spread Spectrum Cellular System,* TIA/EIA Interim Standard-95, July 1993.

[3] 3GPP2, "cdma2000 Standard for Spread Spectrum Systems, Revision D," Feb. 2004.

[4] H. Holma and A. Toskala, *WCDMA for UMTS: HSPA Evolution and LTE*, John Wiley and Sons, 2010.

[5] IEEE Std. 802. 11, Part 11: Wireless LAN Medium Access Contriol (MAC) and Physical Layer (PHY) Specifications, 2007.

[6] D. Sarwate and M. Pursley, "Crosscorrelation properties of pseudo-random and related sequences," *Proc. IEEE*, vol. 68, pp. 593–619, May 1980.

[7] J. S. Lee and L. E. Miller, *CDMA systems Engineering Handbook*, Artech House, Jan. 1998.

[8] S. Tamura, S. Nakano, and K. Okazaki, "Optical code-multiplex transmission by gold sequences," *J. Lightw. Technol.*, vol. 3, no. 1, pp.121–127, Feb. 1985.

[9] Y. H. Lee and S. J. Kim, "Sequence acquisition of DS-CDMA systems employing gold sequences," *IEEE Trans. Veh. Technol.*, vol. 49, pp. 2397–2404, Nov. 2000.

[10] F. Hemmati and D. Schilling, "Upper Bounds on the Partial Correlation of PN Sequences," *IEEE Trans. Commun.*, vol.31, no.7, pp. 917- 922, Jul. 1983.

[11] J. H. Lindholm, "An analysis of the pseudo-randomness properties of subsequences of long m-sequences," *IEEE trans. Inform. Theory*, vol. IT-14, pp.569-576, Jul. 1968.

[12] N. E. Berkir, R. A. Scholtz, and L. R. Welch, "Partial-period correlation properties of PN sequence," in *1978 National Telecommunication Conf. Rec.*, vol. 3, pp. 35.1.1-35.1.4, Dec. 1978.

[13] J. Yang, D. Bao, and M. Ali, "PN offset planning in IS-95 based CDMA systems," in *Proc. IEEE 47th Veh. Technol. Conf.*, May 1997, pp.1435-1439.

[14] D. I. Kim, E. Hossain, and V. K. Bhargava, "Dynamic random access code assignment for prioritized packet data transmission in WCDMA networks," *IEEE Wireless Commun.*, vol.2, no.5, pp. 911- 925, Sep. 2003.

[15] W. H. Sheen, C. C. Tseng, and J. S. Ho, "Burst synchronization of slotted random access with preamble power ramping in the reverse link of CDMA systems," *IEEE Wireless Commun.*, vol.2, no.5, pp. 953- 963, Sep. 2003.

[16] D. Chu, " Polyphase codes with good periodic correlation properties (Corresp.)," *IEEE Trans. Inform. Theory*, vol.18, no.4, pp. 531-532, Jul. 1972.

[17] 3rd Generation Partnership Project Technical Specification Group Radio Access Network Evolved Universal Terrestrial Radio Access (E-UTRA) Physical Channels and Modulation (Release 11), TS 36.211 ver.11.0.0, Sep. 2012.

[18] J. W. Kang, Y. Whang, B. H. Ko, and K. S. Kim, "Generalized Cross-Correlation Properties of Chu Sequences," *IEEE Trans. Inform. Theory*, vol.58, no.1, pp.438-444, Jan. 2012.

[19] S. Sesia, I. Toufik and M. Baker, *LTE-The UMTS Long Term Evolution: From Theory to Practice,* John Wiley and Sons, 2009.

[20] M. Wang, L. Xiao, T. Brown, and M. Dong, "Optimal Symbol Timing for OFDM Wireless Communications," *IEEE Trans. Wireless Commun.*, vol. 8, no. 10, Oct. 2009.

[21] En Zhou, Yuyu Yan, and Wenbo Wang, "A Novel Timing Synchronization Method for Localized OFDMA Uplink System," *IEEE International Commun. Conf (ICC)*, vol. 11, pp. 5086-5090, June 2006.

[22] K. Chang, W. Y. Lee, and H. K. Chung, "Frequency-immune and low-complexity symbol timing synchronization scheme in OFDM systems," in *Proc. IEEE PIMRC*, pp. 922-926, Sep. 2010.

[23] Y. Wen, W. Huang, and Z. Zhang, "CAZAC sequence and its application in LTE random access," in *Proc. IEEE Inf. Theory Workshop (ITW)*, Oct. 2006, pp.544-547.

[24] R. Frank, S. Zadoff, and R. Heimiller, "Phase shift pulse codes with good periodic correlation properties (Corresp.)," *IRE Trans. Inform. Theory*, vol.8, no.6, pp.381-382, Oct. 1962.

[25] R. Heimiller, "Phase shift pulse codes with good periodic correlation properties," *IRE Trans. Inform. Theory*, vol.7, no.4, pp.254-257, Oct. 1961.

[26] Y. Zhou, J. Wang, and M. Sawahashi, "Downlink transmission of broadband OFCDM Systems-part II: effect of Doppler shift," *IEEE Trans. Commun.*, vol.54, no.6, pp.1097-1108, June. 2006.

[27] 3rd Generation Partnership Project Technical Specification Group Radio Access Network Evolved Universal Terrestrial Radio Access (E-UTRA) User Equipment (UE) radio transmission and reception (Release 11), TS 36.101 ver.11.2.0, Sep. 2012.

[28] M. Gul, S. Lee, and X. Ma, "Robust synchronization for OFDM employing Zadoff-Chu sequence", *the 46th Annual Conference on Information Science and Systems (CISS)*, pp.1-6, Mar. 2012.

[29] Min Hua, Michael Mao Wang, Wenjie Yang, Xiaohu You, Feng Shu, Jianxin Wang, Weixing Sheng, and Qian Chen, "Analysis of the Frequency Offset Effect on Random Access Signals", *IEEE Trans. Commun*, vol. 61, no. 11, pp.4728-4740, Nov. 2013.